\documentclass[aps,twocolumn,superscriptaddress,showpacs]{revtex4}

\usepackage[dvips]{graphicx}
\usepackage{color}  % for color  usage: \textcolor{red}{text}
\usepackage{amssymb}
\newcommand{\nc}{\newcommand}           % new command
\nc{\vc}[1]     {\mbox{\boldmath $#1$}} % boldmath(vector)
\nc{\mapleft}[1]{                       % something under arrow
 \smash{\mathop{                      %
  \hbox to 0.90cm{\rightarrowfill} }\limits_{#1}}}

\nc{\beq}     {\begin{eqnarray}}
\nc{\eeq}    {\end{eqnarray}}

\nc{\red}[1]    {\textcolor{red}{#1}}  % red
\nc{\blue}[1]   {\textcolor{blue}{#1}}  % blue
\nc{\green}[1]   {\textcolor{green}{#1}}  % green

\nc{\mydraft}	{\setlength{\topmargin}{-1.5cm}}
\mydraft

\begin{document}

\title{Successive variational method of the tensor-optimized antisymmetrized molecular dynamics for central interaction in finite nuclei}

\author{Takayuki Myo\footnote{takayuki.myo@oit.ac.jp}}
\affiliation{General Education, Faculty of Engineering, Osaka Institute of Technology, Osaka, Osaka 535-8585, Japan}
\affiliation{Research Center for Nuclear Physics (RCNP), Osaka University, Ibaraki, Osaka 567-0047, Japan}

\author{Hiroshi Toki\footnote{toki@rcnp.osaka-u.ac.jp}}
\affiliation{Research Center for Nuclear Physics (RCNP), Osaka University, Ibaraki, Osaka 567-0047, Japan}

\author{Kiyomi Ikeda\footnote{k-ikeda@postman.riken.go.jp}}
\affiliation{RIKEN Nishina Center, Wako, Saitama 351-0198, Japan}

\author{Hisashi Horiuchi\footnote{horiuchi@rcnp.osaka-u.ac.jp}}
\affiliation{Research Center for Nuclear Physics (RCNP), Osaka University, Ibaraki, Osaka 567-0047, Japan}

\author{Tadahiro Suhara\footnote{suhara@matsue-ct.ac.jp}}
\affiliation{Matsue College of Technology, Matsue 690-8518, Japan}

\date{\today}

\begin{abstract}%
Tensor-optimized antisymmetrized molecular dynamics (TOAMD) is the basis of the successive variational method for the nuclear many-body problem.
We apply TOAMD to finite nuclei described by the central interaction with strong short-range repulsion, and compare the results with those from the unitary correlation operator method (UCOM).
In TOAMD, the pair-type correlation functions and their multiple products are operated to the antisymmetrized molecular dynamics (AMD) wave function. 
We show the results of TOAMD using the Malfliet-Tjon central potential containing the strong short-range repulsion. 
By adding the double products of the correlation functions in TOAMD, the binding energies are converged quickly to the exact values 
of the few-body calculations for $s$-shell nuclei.
This indicates the high efficiency of TOAMD for treating the short-range repulsion in nuclei.
We also employ the $s$-wave configurations of nuclei with the central part of UCOM, which reduces the short-range relative amplitudes of nucleon pair in nuclei to avoid the short-range repulsion.
In UCOM, we further perform the superposition of the $s$-wave configurations with various size parameters, which provides a satisfactory solution of energies close to the exact and TOAMD values.
\end{abstract}

\maketitle 

%%%%%%%%%%%%%%%%%%%%%%%%%%%%%%%%%%%%%%%%
\section{Introduction} \label{sec:intro}

The nucleon-nucleon ($NN$) interaction plays an essential role in determining the nuclear structure.
The $NN$ interaction has a strong repulsion at short distance in addition to a strong tensor force at long and intermediate distances \cite{pieper01,wiringa95}.
These two characteristics of the $NN$ interaction provide the high-momentum components of nucleon motion in nuclei, which should be treated in the nuclear wave function. 
The short-range repulsion reduces the relative amplitudes of a nucleon pair at short-distance in a nucleus.
The tensor force produces the characteristic $D$-wave state of a nucleon pair in nuclei, which comes from the strong $S$-$D$ coupling of the tensor force. 
This $D$-wave state is spatially compact as compared with the $S$-wave state due to the high-momentum component of the tensor correlation \cite{ikeda10,ong13}. 

So far, we have described the short-range and tensor correlations in nuclei with two-kinds of theoretical methods.
One is the shell model-type approach, which we call the tensor-optimized shell model (TOSM) \cite{myo05,myo09,myo11}.
In TOSM, we fully optimize the two-particle two-hole (2p2h) states in the wave function without any truncation for the particle states.
These 2p2h excitations can represent the strong tensor correlation in nuclei.
On the other hand, it is difficult to express the short-range correlation for the shell model-type basis states.
Hence, we combine TOSM with the central part of the unitary correlation operator method (central-UCOM) in which the central-type short-range correlation is explicitly treated \cite{feldmeier98,neff04}. 
In the central-UCOM, the central-type shift operator is introduced in order to reduce the short-range amplitudes of the relative motion of a nucleon pair in nuclei.
In UCOM, the transformed Hamiltonian is truncated up to two-body operators, while the exact transformation produces many-body operators.
This two-body approximation of UCOM is considered to be reasonable for the short-range correlation.
In TOSM+UCOM, we treat the tensor and short-range correlations explicitly in the wave function.
This method nicely works to describe the shell model-like states with the correct order of the energy level in the $p$-shell nuclei \cite{myo12,myo14}.

On the other hand, it is important to treat the nuclear clustering phenomena, which are difficult to treat in the shell model-type approach \cite{barrett13}.
The nuclear clustering is an important aspect of the nuclear structure, such as the triple-$\alpha$ Hoyle state in $^{12}$C \cite{ikeda68,horiuchi12}.
Recently, we have developed a new variational theory about the clustering description of nuclei from the $NN$ interaction  \cite{myo15,myo16}.
We employ the antisymmetrized molecular dynamics (AMD) \cite{kanada03,kanada12} as the basis state
and introduce two-kinds of the correlation functions of the tensor-operator type and the central-operator type to treat the $NN$ interaction.
These correlation functions are multiplied to the AMD wave function and superposed with the original AMD wave function.
We name this framework the tensor-optimized antisymmetrized molecular dynamics (TOAMD) \cite{myo15}.
We can extend the scheme of TOAMD by using the series of the multiple products of the correlation functions as the power expansion successively.
In TOAMD, the products of the Hamiltonian and the correlation functions appear, which become the series of many-body operators in the cluster expansion.
We take care of all the resulting many-body operators, which ensures TOAMD as the variational method.
The formulation of TOAMD is general for nuclei with various mass numbers.
In the previous work \cite{myo16}, we have calculated the $s$-shell nuclei with TOAMD within the double products of the correlation functions,
and have shown that TOAMD nicely reproduces the results of Green's function Monte Carlo (GFMC) using the AV8$^\prime$ bare $NN$ interaction \cite{kamada01}.

It is interesting to investigate how TOAMD describes the central-type correlation coming from the central $NN$ interaction,
which has a short-range repulsion, in the same scheme as the previous work \cite{myo16}.
This subject can give a scope to apply TOAMD to other fields.
In this study, we focus on the central-type correlation and use the Malfliet-Tjon V central potential having a Yukawa-type tail and a strong short-range repulsion \cite{malfliet69}.
We take care of the short-range correlation using the central-type correlation functions and demonstrate the results of TOAMD for $s$-shell nuclei.
Furthermore, the central-UCOM could be a powerful method for the treatment of the short-range repulsion.
We shall compare the results of TOAMD with those using the central-UCOM, which makes clear how the central-UCOM describes the short-range correlations of nuclei quantitatively under the two-body approximation of the unitary transformation.

This paper is organized as follows.
In Sec.~\ref{sec:method}, we explain TOAMD and UCOM for the treatment of short-range correlations.
In Sec.~\ref{sec:result}, we present the results of $s$-shell nuclei with the two methods.
A summary is provided in Sec.~\ref{sec:summary}.

%%%%%%%%%%%%%%%%%%%%%%%%%%%%%%%%%%%%%%%%%%%%%%%%%%%%%%%%%%%%%%%%
\section{Hamiltonian and methods}\label{sec:method}

%%%%%%%%%%%%%%%%%%%%%%%%%%%%%%%%%%%%%%%%%%%%%%
\subsection{Hamiltonian}
We use the Hamiltonian with two-body interaction $V$ for a mass number $A$ as
\begin{eqnarray}
    H
&=& T+V
~=~ \sum_i^{A} t_i - T_{\rm cm} + \sum_{i<j}^{A} v_{ij} . 
    \label{eq:Ham}
\end{eqnarray}
Here, $t_i$ and $T_{\rm cm}$  are the kinetic energies of each nucleon and the center-of-mass, respectively.
We aim at investigating how TOAMD and UCOM describe the central-type correlation including the short-range correlation. 
For this purpose, we use the Malfliet-Tjon V (MT-V) central potential as an $NN$ interaction \cite{malfliet69}, 
in the following form with a Yukawa-type tail and a strong short-range repulsion:
\begin{eqnarray}
   v(r)
&=& 1458.05\ \frac{e^{-3.11 r}}{r} - 578.09\ \frac{e^{-1.55r}}{r},
   \label{eq:MT-V}
\end{eqnarray}
in units of MeV and $r$ in units of fm. We also take $\hbar^2/m=41.47$ MeV fm$^2$ \cite{zabolitzky82}.

%%%%%%%%%%%%%%%%%%%%%%%%%%%%%%%%%%%%%%%%%%%%%%
\subsection{Tensor-optimized antisymmetrizd molecular dynamics (TOAMD)} \label{sec:TOAMD}

We explain the essential formulation of TOAMD. The details of TOAMD are given in Ref. \cite{myo15}.
We start from the AMD wave function, which is the Slater determinant consisting of the Gaussian wave packets of nucleons.
The AMD wave function $\Phi_{\rm AMD}$ is defined as
\begin{eqnarray}
\Phi_{\rm AMD}
&=& \frac{1}{\sqrt{A!}} {\rm det} \left\{ \prod_{i=1}^A \phi_i \right\}~,
\label{eq:AMD}
\\
\phi(\vec r)&=&\left(\frac{2\nu}{\pi}\right)^{3/4} e^{-\nu(\vec r-\vec D)^2} \chi_{\sigma} \chi_{\tau}~.
\label{eq:Gauss}
\end{eqnarray}
The single-nucleon wave function $\phi(\vec r)$ consists of a Gaussian wave packet with a range parameter $\nu$ and a centroid position $\vec D$, 
the spin part $\chi_{\sigma}$ and isospin part $\chi_{\tau}$.
In this study of $s$-shell nuclei, $\chi_{\sigma}$ is fixed with the up or down component and $\chi_{\tau}$ is a proton or neutron.
The range parameter $\nu$ is common for all nucleons and this condition factorizes the center-of-mass wave function from $\Phi_{\rm AMD}$.

In the usual TOAMD we include two-kinds of correlations induced by the tensor force and short-range repulsion.
Following the concept given in Refs. \cite{sugie57,nagata59},
we introduce the pair-type correlation functions $F_D$ for tensor force and $F_S$ for short-range repulsion and multiply them to the AMD wave function, individually.
We superpose these components with the AMD wave function. The correlation functions $F_D$ and $F_S$ are determined variationally.
This concept of TOAMD is motivated from the success of TOSM \cite{myo09,myo14}.

In this study, we focus on the short-range repulsion of the central force and consider only the short-range correlation function $F_S$ in TOAMD.
We define the TOAMD wave function for short-range correlation as
\begin{eqnarray}
\Phi_{\rm TOAMD}^{\rm single}
&=& (1+F_S) \times\Phi_{\rm AMD}~,
\label{eq:TOAMD1}
\\
F_S
&=& \sum_{t=0}^1\sum_{s=0}^1\sum_{i<j}^A f^{t,s}_{S}(r_{ij})\,O^t_{ij}\,O^s_{ij}~,
\label{eq:Fs}
\end{eqnarray}
with relative coordinate $\vec r_{ij}=\vec r_i - \vec r_j$, $O^t_{ij}=(\vec \tau_i\cdot \vec \tau_j)^t$ and $O^s_{ij}=(\vec\sigma_i\cdot \vec\sigma_j)^s$.
Here $t$ and $s$ represent the isospin and spin channel of a pair, respectively.
We call this form of the TOAMD wave function with a single correlation function $F_S$ ``single TOAMD.''
The function $F_S$ affects the relative motion of a nucleon pair in $\Phi_{\rm AMD}$.
This $F_S$ is a scalar operator and does not change the angular-momentum state of $\Phi_{\rm AMD}$.

As an extension of Eq.~(\ref{eq:TOAMD1}), we increase the order of correlation function in TOAMD by adding double products of $F_S$ according to the previous study \cite{myo15}.
Similar to the single case, we define the TOAMD wave function with the double correlation functions below, which is called ``double TOAMD'': 
\begin{eqnarray}
\Phi_{\rm TOAMD}^{\rm double}
&=& (1+F_S+F_S F_S) \times\Phi_{\rm AMD}~.
\label{eq:TOAMD2}
\end{eqnarray}
This form is the power series expansion in terms of the correlation function $F_S$.
It is noted that $F_S$ in each term in Eq.\,(\ref{eq:TOAMD2}) are independent and variationally determined.
Hence, we have three-kinds of $F_S$ in the double TOAMD.
This treatment of the correlation functions increases the variational degrees of freedom in TOAMD successively.
For simplicity, we denote the symbol of $F_S$ commonly in $\Phi_{\rm TOAMD}^{\rm double}$.

Physically, the correlation function $F_S$ can excite two nucleons in the AMD state to the high-momentum region.
This corresponds to the 2p2h excitations in the shell model picture. 
The $F_S F_S$ term in Eq.~(\ref{eq:TOAMD2}) is the situation in which the short-range correlations can occur twice at the same time in a nucleus, corresponding to the 4p4h excitations at maximum.
This formulation of TOAMD is independent of the mass number.
We can also increase the power expansion of $F_S$ successively in TOAMD to the next order such as $F_SF_SF_S$ in case we want to increase the accuracy of the numerical results.
We can examine the convergence of the solutions with the power of the correlation functions.

The total energy in TOAMD is given as
\begin{eqnarray}
E_{\rm TOAMD}
&=&\frac{\langle\Phi_{\rm TOAMD} |H|\Phi_{\rm TOAMD}\rangle}{\langle\Phi_{\rm TOAMD} |\Phi_{\rm TOAMD}\rangle}\nonumber\\
&=&\frac{\langle\Phi_{\rm AMD} | \tilde{H} |\Phi_{\rm AMD}\rangle}{\langle\Phi_{\rm AMD} | \tilde{N} |\Phi_{\rm AMD}\rangle},
\label{eq:E_TOAMD}
\end{eqnarray}
where $\tilde{H}$ and $\tilde{N}$ are the correlated Hamiltonian and norm, respectively.
We calculate the matrix elements of these operators with the AMD wave function.
The correlated operators $\tilde{H}$ and $\tilde{N}$ include the products of correlation functions, such as $F_S^\dagger H F_S$ and $F_S^\dagger F_S$, respectively. 
These operators are decomposed into the series of many-body operators.
In the case of the kinetic energy $T$, $F_S^\dagger T F_S$ is expanded to from two-body to five-body operators with various combinations of particle index.
For the two-body interaction $V$, $F_S^\dagger V F_S$ is expanded to from two-body to six-body operators. 
In the same way, $F_S^\dagger F_S^\dagger V F_S F_S$ gives ten-body operators at maximum.
We classify these many-body operators in terms of the cluster expansion method, the detailed procedure of which is given in Ref.~\cite{myo15}.
We employ all the resulting many-body operators without any truncation and calculate the matrix elements of each operator with the AMD wave function.
The procedure is performed systematically for any order of the multiple products of the correlation functions in TOAMD.
In general, many-body operators produce larger number of terms of the cluster expansion for larger mass systems and 
the calculation of their matrix elements needs much more computational cost.
In this study, we discuss the $s$-shell nuclei with up to the mass number of four. Hence, at most four-body operators are treated.

The present TOAMD wave function has two-kinds of variational functions, the AMD wave function $\Phi_{\rm AMD}$ and the correlation functions $F_S$.
We determine these functions using the Ritz variational principle with respect to the TOAMD energy as $\delta E_{\rm TOAMD}=0$.
For $\Phi_{\rm AMD}$, the centroid positions of the Gaussian wave packets $\{\vec D_i\}$ ($i=1,\cdots,A$) in Eq.~(\ref{eq:Gauss}) are determined variationally
by using the cooling method \cite{kanada03}.

The radial forms of $F_S$ are optimized in four spin-isospin channels to minimize the energy $E_{\rm TOAMD}$ in Eq.~(\ref{eq:E_TOAMD}).
We adopt the Gaussian expansion method to express the pair functions $f^{t,s}_{S}(r)$ in Eq.~(\ref{eq:Fs}) given as
\begin{eqnarray}
   f^{t,s}_S(r)
&=& \sum_{n=1}^{N_G} C^{t,s}_n\, e^{-a^{t,s}_n r^2}~.
   \label{eq:cr_S}
\end{eqnarray}
Here, $C^{t,s}_n$ and $a^{t,s}_n$ are the variational parameters.
We take the Gaussian basis number $N_G=10$ until the solutions are converged.
For the Gaussian ranges $a^{t,s}_n$, we search for the optimized values from short to long ranges to express the spatial correlation adequately.
The coefficients $C^{t,s}_n$ are the linear parameters in the TOAMD wave function and determined variationally by diagonalizing the Hamiltonian matrix elements.
For the $F_SF_S$ term in Eq.~(\ref{eq:TOAMD2}), the products of two Gaussian functions in Eq. (\ref{eq:cr_S}) become the basis functions and 
correspondingly the products of $C^{t,s}_n$ are the variational parameters.

In the calculation of Hamiltonian matrix elements, we express the $NN$ interaction as a sum of Gaussians.
In order to calculate the matrix elements of many-body operators in $\tilde{H}$ and $\tilde{N}$, we use the Fourier transformation of the Gaussians in the correlation functions $F_S$ and the $NN$ interaction \cite{myo15,goto79}.
This transformation decomposes the many-body operators containing various sets of the relative coordinates in the exponent into the separable form of the single particle coordinates with the plane wave form.
In the momentum space, the matrix elements of the many-body operators result in the products of single-particle matrix elements of the plane wave.
Using the single-particle matrix elements in AMD, we perform the multiple integration of the associated momenta and obtain the matrix elements of TOAMD.

%%%%%%%%%%%%%%%%%%%%%%%%%%%%%%%%%%%%%%%%%%%%%%
\subsection{Unitary correlation operator method (UCOM)}
We explain the central-UCOM for the short-range central correlation \cite{feldmeier98,neff04}. 
One introduces the following unitary operator $C$
\begin{eqnarray}
C     &=&\exp(-i\sum_{i<j} g_{ij})~.
\label{eq:ucom}
\end{eqnarray}
We express the wave function $\Psi$ including the short-range correlation in terms of the uncorrelated wave function $\Phi$ as $\Psi=C\Phi$. 
The transformed Schr\"odinger equation is given as $\hat H \Phi=E\Phi$ where the transformed Hamiltonian has the relation of $\hat H=C^\dagger H C$. 
In principle, the operator $C$ becomes a many-body one because of the two-body operator in the exponent. 
Similarly, the transformed Hamiltonian $\hat H$ becomes the series of many-body operators.
In the case of the short-range correlation, it seems reasonable to truncate $\hat H$ at the two-body level \cite{feldmeier98}.

Two-body Hermite operator $g$ in Eq.~(\ref{eq:ucom}) is defined as
\begin{eqnarray}
g &=& \frac12 \left\{ p_r s(r)+s(r)p_r\right\} ~,
\label{eq:ucom_g}
\end{eqnarray}
\begin{eqnarray}
\frac{dR_+(r)}{dr}&=& \frac{s [ R_+(r) ]}{s(r)} ~, 
\end{eqnarray}
where the operator $p_r$ is the radial component of the relative momentum, conjugate to the relative distance $r$. 
The function $s(r)$ represents the amount of the shift of the relative wave function at distance $r$. 
In the central-UCOM, the function $R_+(r)$ is often used instead of $s(r)$, where $R_+(r)$ is the transformed distance of the original distance $r$. 
This $R_+(r)$ plays a role to reduce the short-range amplitude of the relative wave functions to avoid the short-range repulsion in the $NN$ interaction. 
The explicit transformations of the various operators are given in Refs. \cite{feldmeier98,neff04}. 

In many-body case, the shift operator $g$ in Eq.~(\ref{eq:ucom_g}) is introduced for every nucleon pair in nuclei. 
The amount of the shifts, namely the function $R_+(r)$ is determined in the energy minimization of the total system
under the two-body approximation of the UCOM transformation.
We parametrize $R_+(r)$ in the same form proposed by Feldmeier and Neff \cite{feldmeier98,neff04}.
\begin{eqnarray}
	R_+(r)
&=&	r + \alpha \left(\frac{r}{\beta}\right)^\gamma \exp[-\exp(\frac{r}{\beta})] , 
        \label{eq:R+}
\end{eqnarray}
where $\alpha$, $\beta$, and $\gamma$ are the variational parameters.

\begin{figure}[t]
\begin{center}
\includegraphics[width=7.8cm,clip]{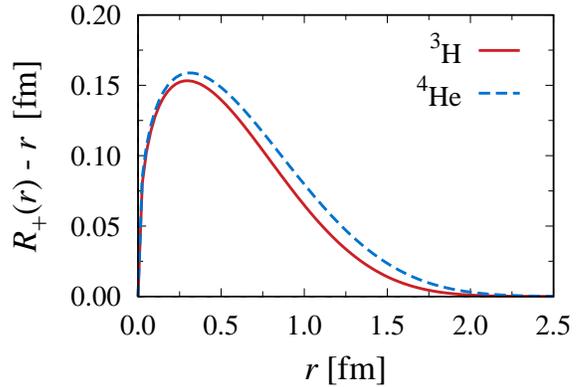}
\caption{The functions $R_+(r)$ in $^3$H (solid) and $^4$He (dashed), respectively. 
The differences between $R_+(r)$ and the original distance $r$ are shown as the amount of shift of the wave function.}
\label{fig:R+}
\end{center}
\end{figure}

In the calculation with UCOM, we choose the uncorrelated wave function $\Phi$ as the $(0s)^A$ configurations of the harmonic oscillator (HO) 
wave functions with the length parameter $b$ for $^3$H and $^4$He, respectively. 
The length $b$ has a relation with $\nu$ in AMD in Eq.~(\ref{eq:Gauss}) when $\vec D=0$, as 
\begin{eqnarray}
 b&=& \frac{1}{\sqrt{2\nu}}.
\end{eqnarray}
The correlated wave function also depends on the length $b$ as $\Psi(b)=C\Phi(b)$.
We determine the length $b$ and $\alpha$, $\beta$, and $\gamma$ in $R_+(r)$ in the energy minimization of each nucleus using MT-V.
In Table \ref{tab:R+}, we list four parameters obtained for $^3$H and $^4$He, respectively.
In Fig.~\ref{fig:R+}, the distribution of $R_+(r)-r$ is shown as the difference between the transformed and the original distances of a nucleon pair, 
which represents the amount of shift of the relative wave function at $r$.
It shows a maximum shift of about 0.15 fm at about 0.3 fm of the relative distance in each nucleus.

%%%%%%%%%%%%%%%%%%%%%%%%%%%%%% 
% Src2.6/Data_MTV
\begin{table}[t]
\begin{center}
\caption{The length $b$ in units of fm [$\nu=(2b^2)^{-1}$ in units of fm$^{-2}$] and three parameters in $R_+(r)$ of UCOM 
 for $^3$H and $^4$He with MT-V.}
\label{tab:R+} 
\begin{tabular}{c|cccc}
\noalign{\hrule height 0.5pt}
              & $b$~~~~~($\nu$) & $\alpha$ &  $\beta$ & $\gamma$ \\
\noalign{\hrule height 0.5pt}
$^3$H         &~1.44~ (0.24)~~&~ 0.944~  &~ 1.018~  & ~0.389~ \\
$^4$He        &~1.23~ (0.33)~~&~ 0.954~  &~ 1.121~  & ~0.369~ \\
\noalign{\hrule height 0.5pt}
\end{tabular}
\end{center}
\end{table}
%%%%%%%%%%%%%%%%%%%%%%%%%%%%%%
%%%%%%%%%%%%%%%%%%%%%%%%%%%%%%
%%%%%%%%%%%%%%%%%%%%%%%%%%%%%%

In the central-UCOM, the short-range correlation is included in the wave function, but the long-range and intermediate-range correlations do not change.
In order to improve the $s$-wave HO configuration $\Phi(b)$, we superpose the basis states with various length parameters $b$.
This corresponds to the generator coordinate method (GCM).
We use the common UCOM transformation $R_+(r)$ as is shown in Table \ref{tab:R+} for these basis states. 
The total GCM wave function is expressed in the linear combination of the internal wave functions of the correlated basis states $\Psi_{\rm int}(b)$ as
\begin{eqnarray}
\Psi_{\rm GCM}  &=& \sum_{i}^N c_i \Psi_{\rm int}(b_i).
\end{eqnarray}
The internal wave function is defined as 
\begin{eqnarray}
\Psi(b)&=& \Psi_{\rm int}(b) \times \left(\frac{A}{\pi b^2} \right)^{3/4} e^{-A (\vec{r}_{\rm G}/b)^2/2}.
\end{eqnarray}
The correlated basis state $\Psi(b)$ has the $0s$ center-of-mass state with the coordinate $\vec{r}_{\rm G}$, which is factorized and independent of UCOM.
The coefficients $\{c_i\}$ are determined by the diagonalization of the Hamiltonian matrix elements with UCOM. 
We take the basis number $N$ as 40 covering a wide range of $b$ to get the converged solutions.

%%%%%%%%%%%%%%%%%%%%%%%%
\section{Results}\label{sec:result}

\subsection{$^3$H}
We show the results of $^3$H with MT-V.
We first demonstrate the calculation of TOAMD with the single correlation function, namely the single TOAMD, $\Phi_{\rm TOAMD}^{\rm single}$ in Eq.~(\ref{eq:TOAMD1}).
This is done to investigate the configuration of AMD wave function $\Phi_{\rm AMD}$.
In the energy variation of the TOAMD wave function, we optimize the range parameter $\nu$ and the centroid positions of nucleons $\vec D$ in $\Phi_{\rm AMD}$.
The energy minimum is obtained as $-7.68$ MeV with $\nu$=0.20 fm$^{-2}$ and $\vec D=0$ for all nucleons,
which is equivalent to the shell-model states of $(0s)^3$.
This result indicates that the $s$-wave configuration is preferred in AMD even with the short-range repulsion. 
The same result of $\vec D=0$ is obtained for $^3$H and $^4$He using the AV8$^\prime$ bare $NN$ interaction having the tensor and $LS$ components \cite{myo16}.
In the present analysis, we keep this condition of $\vec D=0$ in TOAMD for $^3$H and $^4$He.

%%%%%%%%%%%%%%%%%%%%%%%%%%%%%%%%%%%%
\begin{figure}[t]
\centering
\includegraphics[width=8.3cm,clip]{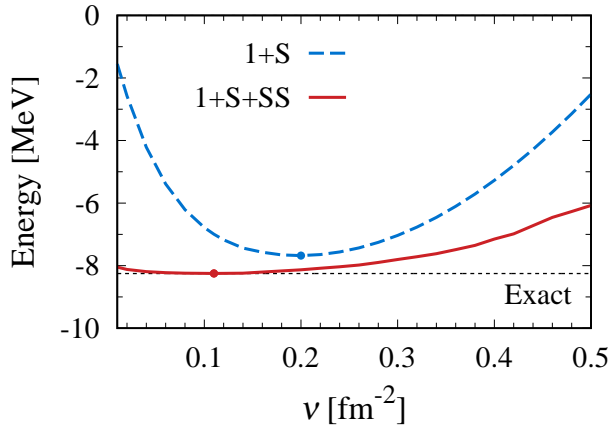}
\caption{Energy surface of $^3$H as function of range parameter $\nu$ in TOAMD with single (dashed line) and double (solid line) correlation functions. Solid circles are the energy minimum points. Dotted line represents the exact energy in the few-body calculations \cite{varga95}.}
\label{fig:ene_3H}
\end{figure}
%%%%%%%%%%%%%%%%%%%%%%%%%%%%%%%%%%%%
%%%%%%%%%%%%%%%%%%%%%%%%%%%%%%%%%%%%
\begin{figure}[t]
\centering
\includegraphics[width=8.3cm,clip]{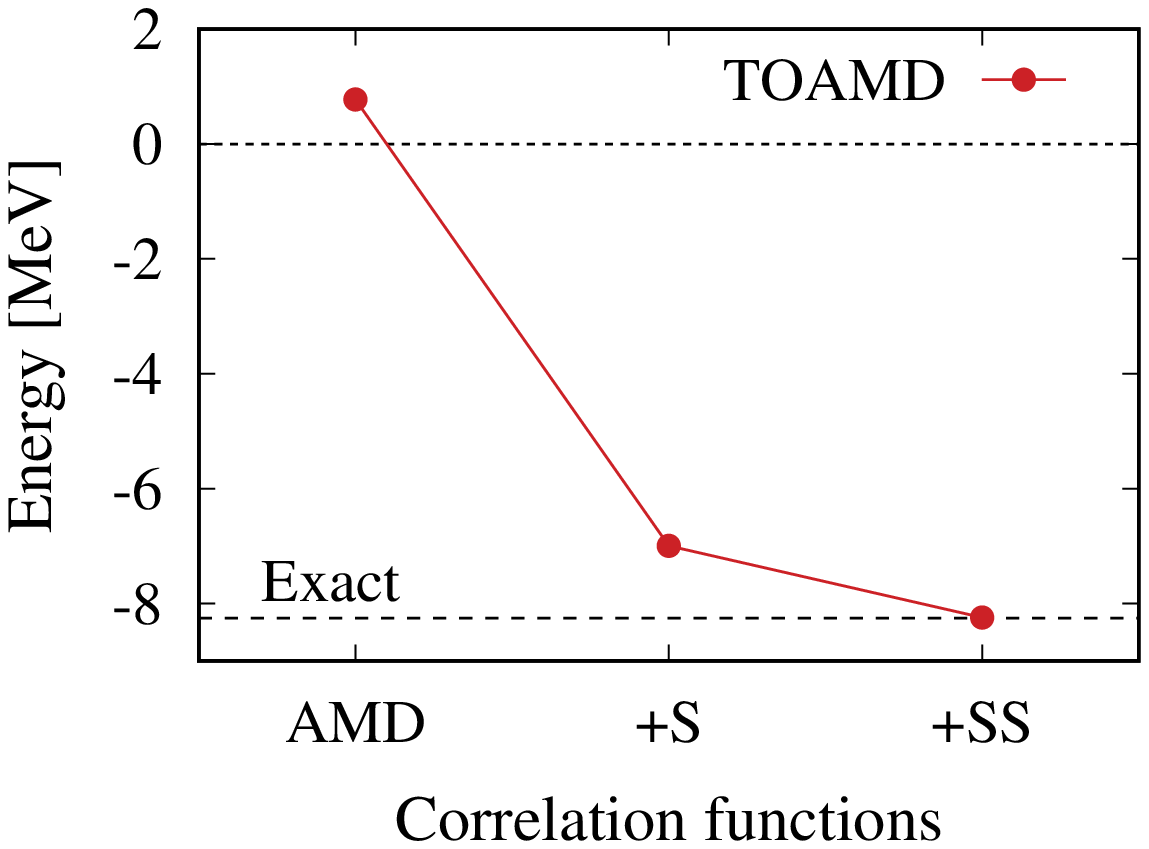}
\caption{Convergence of the energy of $^3$H with respect to the order of TOAMD. The range parameter $\nu$ is fixed as 0.11 fm$^{-2}$ which is determined in the double TOAMD.}
\label{fig:ene_3H_cnv}
\end{figure}
%%%%%%%%%%%%%%%%%%%%%%%%%%%%%%%%%%%%
%%%%%%%%%%%%%%%%%%%%%%%%%%%%%% 
% Src2.6/Data_MTV
\begin{table}[t]
\begin{center}
\caption{Energies of $^3$H($\frac12^+$) and $^4$He ($0^+$) with MT-V potential in TOAMD in units of MeV in comparison with other theories.
In UCOM, a two-body approximation is adopted for the transformed Hamiltonian, which produces the slight overbinding of the energy of $^4$He with respect to the few-body calculations. For details, see the text.}
\label{tab:energy} 
\begin{tabular}{c|ccccc}
\noalign{\hrule height 0.5pt}
        &  Exact          & \multicolumn{2}{c}{TOAMD} & \multicolumn{2}{c}{UCOM}\\
        &  \cite{varga95} &~~~Single~&~~Double~~~     & HO           & GCM      \\
\noalign{\hrule height 0.5pt}
$^3$H   &~$-8.25$~~       &~$-7.68$  &~$-8.24$ ~~     &~$-6.51$      &~$-7.91$  \\
$^4$He  &~$-31.36$~~      & $-29.35$ & $-31.28$~~     & $-30.77$     & $-31.73$ \\
\noalign{\hrule height 0.5pt}
\end{tabular}
\end{center}
\end{table}
%%%%%%%%%%%%%%%%%%%%%%%%%%%%%%

In Fig.\,\ref{fig:ene_3H}, we show the energy surface of $^3$H as function of the range $\nu$ in AMD. 
It is noted that the range $\nu$ has a relation of $\hbar^2 \nu/ m = \hbar\omega/2$ and the $\nu$-independence corresponds to the 
$\hbar\omega$-independence in the shell model prescription.
We simply write $F_S$ with the symbol of $S$.
At each value of $\nu$, the correlation functions $F_S$ are optimized. We can clearly confirm the energy minimum in $^3$H in the single TOAMD, denoted by $1+S$.
Second, we increase the order of TOAMD to the double TOAMD $\Phi_{\rm TOAMD}^{\rm double}$ containing up to the $F_S F_S$ term.
In the double TOAMD as is shown in Fig.\,\ref{fig:ene_3H}, denoted by $1+S+SS$, the energy curve becomes deeper at any value of $\nu$ due to the additional $F_S F_S$ term.
The lowest energy is obtained as $-8.24$ MeV with $\nu=0.11$ fm$^{-2}$. 
This energy is very close to the exact value of $-8.25$ MeV reported in the few-body calculations \cite{varga95}.
It is found that the energy curve has a flat region with respect to the variation of $\nu$.
This behavior is characteristic and indicates that the TOAMD wave function can be optimized variationally in the wide range of $\nu$ value.
This represents the flexibility of the correlation functions $F_S$.

In Fig.\,\ref{fig:ene_3H_cnv}, we show the convergence of the energy in the order of TOAMD starting from AMD without the short-range correlation,
where $\nu$ is chosen as 0.11 fm$^{-2}$ optimized in the double TOAMD.
In the figure, +S and +SS mean the single and double TOAMD, respectively.
It is confirmed that the TOAMD energy is converged rapidly.
We also summarize the results in Table \ref{tab:energy}.
At energy minimum, the radius of $^3$H is 1.67 fm in TOAMD while few-body calculations provide 1.68 fm \cite{varga95}.
From these results, TOAMD with double correlation functions agrees with the results of few-body calculations very well. 
This fact indicates that the power series expansion of the correlation function in TOAMD is a powerful method.

%%%%%%%%%%%%%%%%%%%%%%%%%%%%%%%%%%%%
\begin{figure}[t]
\centering
\includegraphics[width=8.3cm,clip]{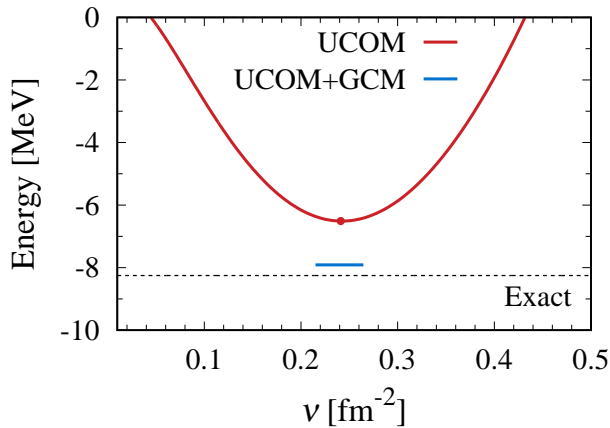}
\caption{Energy surface of $^3$H as function of range parameter $\nu=(2b^2)^{-1}$ in UCOM (solid line) with the result of UCOM+GCM (short solid line).}
\label{fig:ene_3H_UCOM}
\end{figure}
%%%%%%%%%%%%%%%%%%%%%%%%%%%%%%%%%%%%

We compare the results of TOAMD with the UCOM case.
In Fig.\,\ref{fig:ene_3H_UCOM}, we show the energy of $^3$H with UCOM as function of the range parameter $\nu=(2b^2)^{-1}$ with a single $s$-wave HO configuration.
The energy minimum is obtained as $-6.51$ MeV at $\nu=0.24$ fm$^{-2}$ ($b=1.44$ fm) as is listed in Table \ref{tab:energy}.
The result of UCOM underestimate the exact energy by about 1.7 MeV. 
This difference brings the improvement of the uncorrelated $s$-wave HO configuration of $^3$H.
We perform the UCOM+GCM calculation by superposing the basis states with various $\nu$($b$)-parameters.
The UCOM+GCM provides the energy of $-7.91$ MeV and the energy gain is 1.4 MeV from the single configuration case as shown in Fig.~\ref{fig:ene_3H_UCOM}.
The energy becomes close to the exact value and UCOM+GCM fairly reproduces the few-body calculations, but still underestimates the energy slightly by 0.3 MeV.
One of the possible reasons for the small energy difference is that in UCOM+GCM, only the breathing-type correlation is included in the basis function with various spatial sizes.
The other possible reason is the two-body approximation of the UCOM Hamiltonian which corresponds to the 2p2h excitations.
On the other hand, in TOAMD, the correlation function $F_S$ produces the 2p2h excitations and $F_SF_S$ can produce up to the 4p4h excitations. 
This treatment of the particle-hole excitations in TOAMD is beyond the present UCOM+GCM.
In addition, $F_S$ contributes not only to the short-range correlation but also to the long- and intermediate-ranges correlations.

%%%%%%%%%%%%%%%%%%%%%%%%%%%%%%%%%%%%
\begin{figure}[t]
\centering
\includegraphics[width=8.3cm,clip]{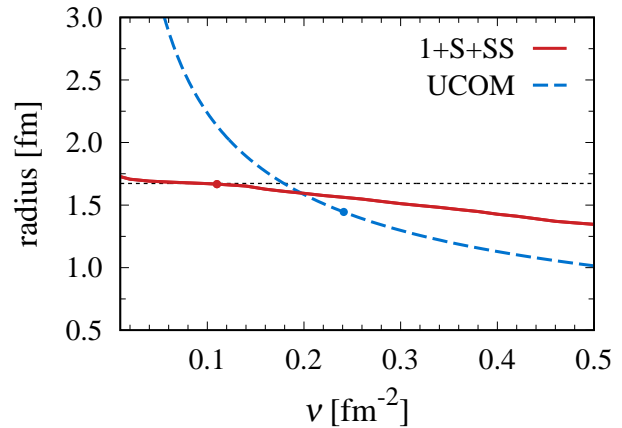}
\caption{Radius of $^3$H as function of range parameter $\nu$ in the double TOAMD (solid line) in comparison with UCOM (dashed line).
Dotted line represents the value (1.67 fm) at the energy minimum in TOAMD.}
\label{fig:radius_3H}
\end{figure}
%%%%%%%%%%%%%%%%%%%%%%%%%%%%%%%%%%%%

In Fig.\,\ref{fig:radius_3H}, the radius of $^3$H in the double TOAMD is shown with respect to the range $\nu$.
The result of UCOM with a single configuration is shown together.
The TOAMD provides flatter curve than the UCOM case.
This flat behavior comes from the flexibility of the correlation functions in TOAMD at various $\nu$ values.
The UCOM calculation with single basis function gives 1.44 fm of radius at the energy minimum, which is small, and the UCOM+GCM calculation gives the larger radius of 1.62 fm, improved due to the effect of GCM.

%%%%%%%%%%%%%%%%%%%%%%%%%%%%%%%%%%%%%%%%%%%%%%%%%%%%%%%%%%%%%%%%%%%%%%%%%%%%%%%%%%%5
\subsection{$^4$He}

%%%%%%%%%%%%%%%%%%%%%%%%%%%%%%%%%%%%
\begin{figure}[t]
\centering
\includegraphics[width=8.3cm,clip]{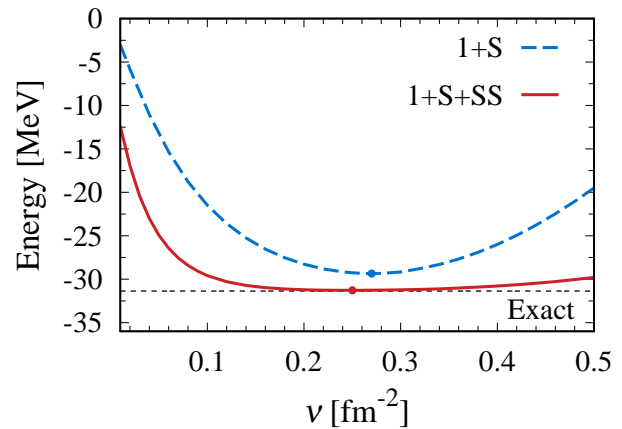}
\caption{Energy surface of $^4$He as function of range parameter $\nu$ in TOAMD with single (dashed line) and double (solid line) correlation functions. Solid circles are the energy minimum points. Dotted line represents the exact energy in the few-body calculations \cite{varga95}.}
\label{fig:ene_4He}
\end{figure}
%%%%%%%%%%%%%%%%%%%%%%%%%%%%%%%%%%%%
%%%%%%%%%%%%%%%%%%%%%%%%%%%%%%%%%%%%
\begin{figure}[t]
\centering
\includegraphics[width=8.3cm,clip]{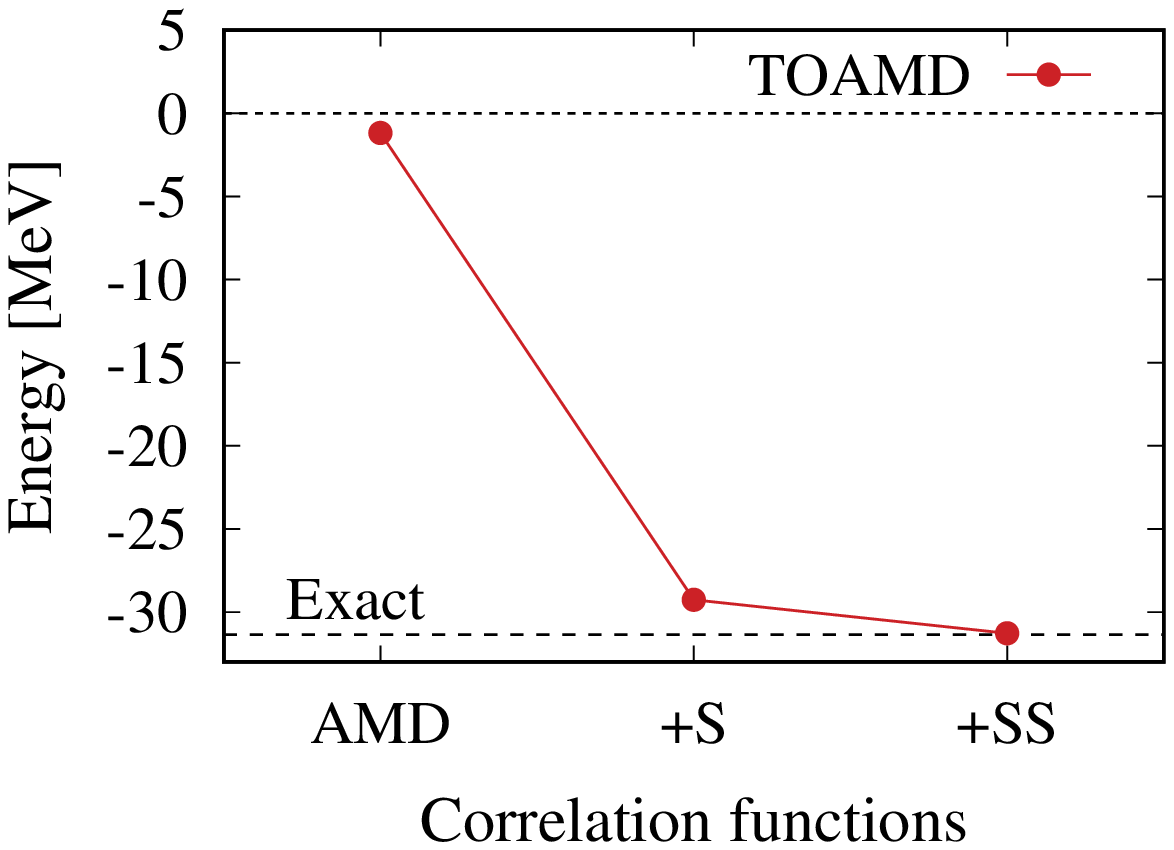}
\caption{Convergence of the energy of $^4$He with respect to the order of TOAMD. The range parameter $\nu$ is fixed as 0.25 fm$^{-2}$ which is determined in the double TOAMD.}
\label{fig:ene_4He_cnv}
\end{figure}
%%%%%%%%%%%%%%%%%%%%%%%%%%%%%%%%%%%%
%%%%%%%%%%%%%%%%%%%%%%%%%%%%%%%%%%%%
\begin{figure}[t]
\centering
\includegraphics[width=8.3cm,clip]{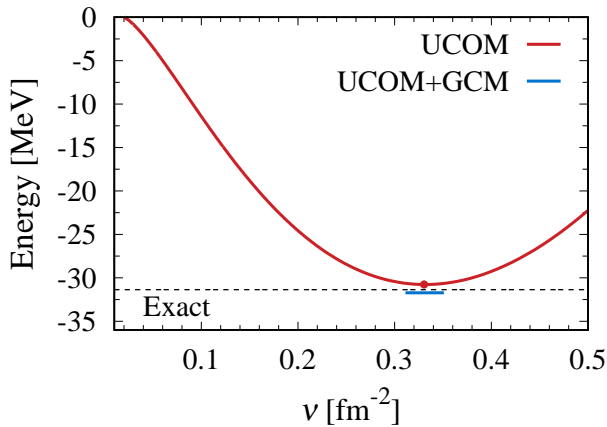}
\caption{Energy surface of $^4$He as function of range parameter $\nu=(2b^2)^{-1}$ in UCOM (solid line) and UCOM+GCM (short solid line).}
\label{fig:ene_4He_UCOM}
\end{figure}
%%%%%%%%%%%%%%%%%%%%%%%%%%%%%%%%%%%%

We show the results of $^4$He. The procedure of the calculation is the same as that of $^3$H.
In the single TOAMD, we confirm that $\vec D=0$ for $^4$He, which is equivalent to the $(0s)^4$ closed configuration.
This result is the same as that using the AV8$^\prime$ bare $NN$ interaction \cite{myo16}.

In Fig.\,\ref{fig:ene_4He}, we show the energy surface of $^4$He as function of the range parameter $\nu$ in AMD.
We show two-kinds of TOAMD calculation with single and double correlation functions, respectively.
In the single TOAMD, the energy minimum is obtained as $-29.35$ MeV with $\nu=0.27$ fm$^{-2}$. 
In the double TOAMD, the energy curve becomes deeper due to the additional $F_SF_S$ term
and the curve has a wide flat region with respect to the variation of $\nu$.
This behavior is common for $^3$H and represents the flexibility of the correlation function $F_S$ in TOAMD.
In the double TOAMD, the energy minimum is obtained as $-31.28$ MeV with $\nu=0.25$ fm$^{-2}$, which is very close to the exact energy of $-31.36$ MeV obtained in the few-body calculations \cite{varga95}.
The radius is 1.40 fm at the energy minimum while GFMC \cite{zabolitzky82} and stochastic variational method \cite{varga95} provide 1.36 fm and 1.41 fm, respectively.

In Fig.\,\ref{fig:ene_4He_cnv}, we show the convergence of the energy of $^4$He in the order of TOAMD with $\nu=0.25$ fm$^{-2}$, similar to the analysis of $^3$H.  
It is confirmed again that the TOAMD energy is converged rapidly.
From these results, TOAMD with double products of the correlation functions agree with the few-body calculations very well for $^4$He in addition to the results of $^3$H.
It is noted that variational Monte Carlo (VMC) calculation gives the energy of $^4$He with $-31.19(5)$ MeV \cite{carlson81},
which is higher than the value of TOAMD ($-31.28$ MeV).
This indicates that TOAMD provides the better solution than that of VMC from the variational point of view.

TOAMD has the advantage in the clustering description based on the AMD basis states. We shall consider the system consisting of several clusters in TOAMD.
For $^4$He, we need the double products of the correlation functions in TOAMD to get the sufficient energy. 
This property indicates that the separated two-$^4$He clusters for $^8$Be naively need the quadruple products of the correlation functions in TOAMD.
It is interesting to investigate the clustering states in TOAMD increasing the power of correlation functions.

%%%%%%%%%%%%%%%%%%%%%%%%%%%%%%%%%%%%
\begin{figure}[t]
\centering
\includegraphics[width=8.5cm,clip]{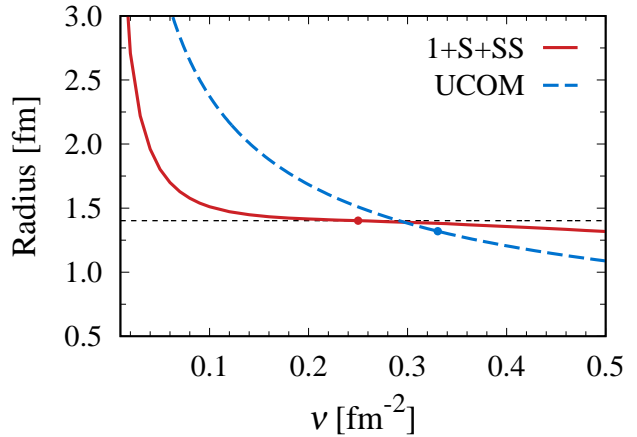}
\caption{Radius of $^4$He as function of range parameter $\nu$ in the double TOAMD (solid line) in comparison with UCOM (dashed line).
Dotted line represents the value (1.40 fm) at the energy minimum state in TOAMD.}
\label{fig:radius_4He}
\end{figure}
%%%%%%%%%%%%%%%%%%%%%%%%%%%%%%%%%%%%

We compare the results of TOAMD with the UCOM case.
The $s$-wave configuration of $^4$He is assumed in UCOM  and we change the range $\nu=(2b^2)^{-1}$ in the HO basis wave function. 
In Fig.\,\ref{fig:ene_4He_UCOM}, the energy minimum is obtained at $\nu=0.33$ fm$^{-2}$ ($b=1.23$ fm) with the energy of $-30.77$ MeV for $^4$He,
as is listed in Table \ref{tab:energy}.
The UCOM calculation slightly underestimates the exact energy by about 0.6 MeV, which is smaller than the $^3$H case of 1.7 MeV. 
This indicates the $(0s)^4$ assumption for $^4$He is rather better than the $(0s)^3$ one for $^3$H.
This is naively considered to come from the strong binding nature of $^4$He.
We further perform the UCOM+GCM calculation by superposing the basis states with various spatial sizes.
The UCOM+GCM provides the energy of $-31.73$ MeV. The energy gain is 1.0 MeV from the single configuration case.
This energy slightly overestimates the exact value of $-31.36$ MeV by about 0.4 MeV. 
One of the possible reasons for the slight overbinding of energy is the missing three-body term of the transformed Hamiltonian in the central-UCOM.
The three-body term of the central-UCOM is not derived so far but its effect is expected to be small physically 
because of the small probability of the three-particle concentration in a small region at the same time.
In this sense, the central-UCOM almost consistently works for the treatment of the short-range correlation in nuclei.

In Fig.\,\ref{fig:radius_4He}, the radius of $^4$He is shown with respect to the range $\nu$.
The double TOAMD provides a rather flatter curve than the UCOM case with a single configuration.
The UCOM calculation gives 1.32 fm for the radius at the energy minimum and the UCOM+GCM calculation make the radius larger, 1.37 fm.

%%%%%%%%%%%%%%%%%%%%%%%
\section{Summary}\label{sec:summary}
We have developed a new variational theory, tensor-optimized antisymmetrized molecular dynamics (TOAMD),  
to describe nuclei from the nucleon-nucleon ($NN$) interaction, in particular, toward the nuclear clustering description.
In TOAMD, the tensor- and central-type correlation functions are multiplied to the AMD wave function 
in order to express the effects of tensor force and short-range repulsion in the $NN$ interaction.
TOAMD is independent of the mass number and extendable by increasing the power of the multiple products of the correlation functions successively.
In TOAMD, the products of the Hamiltonian and the correlation functions produce the many-body operators, which are exactly treated using the cluster expansion. 
This treatment makes TOAMD the variational theory starting from the $NN$ interaction.

In this study, based on the success of TOAMD for $s$-shell nuclei with bare $NN$ interaction \cite{myo16},
we demonstrate the efficiency of TOAMD for the short-range correlation.
We show the results of TOAMD for $s$-shell nuclei with Malfliet-Tjon central potential containing the strong short-range repulsion.
It is found that the TOAMD with double products of the correlation functions nicely reproduces the exact energies in the few-body calculations.
TOAMD with the power series expansion of the correlation functions is expected to be a powerful approach.

We also compare TOAMD with the central part of UCOM (central-UCOM) using the $s$-wave configurations.
In the central-UCOM, the short-range repulsion is almost consistently treated for $s$-shell nuclei
under the two-body approximation of the transformed Hamiltonian.
We also perform the GCM calculation with UCOM to improve the $s$-wave configurations.
The UCOM+GCM provides satisfactory energies, which are comparable to the exact and TOAMD values.
\section*{Acknowledgments}
This work was supported by JSPS KAKENHI Grants No. JP15K05091, No. JP15K17662, and No. JP16K05351.

\section*{References}
%%%%%%%%%%%%%%%%%%%%%%%%%%%%%%%%%%%%%%%%%%%%%%%%%%%%%%%%%%%%%
\def\JL#1#2#3#4{ {{\rm #1}} \textbf{#2}, #4 (#3)}  % Physical Review
\nc{\PR}[3]     {\JL{Phys. Rev.}{#1}{#2}{#3}}
\nc{\PRC}[3]    {\JL{Phys. Rev.~C}{#1}{#2}{#3}}
\nc{\PRA}[3]    {\JL{Phys. Rev.~A}{#1}{#2}{#3}}
\nc{\PRL}[3]    {\JL{Phys. Rev. Lett.}{#1}{#2}{#3}}
\nc{\NP}[3]     {\JL{Nucl. Phys.}{#1}{#2}{#3}}
\nc{\NPA}[3]    {\JL{Nucl. Phys.}{A#1}{#2}{#3}}
\nc{\PL}[3]     {\JL{Phys. Lett.}{#1}{#2}{#3}}
\nc{\PLB}[3]    {\JL{Phys. Lett.~B}{#1}{#2}{#3}}
\nc{\PTP}[3]    {\JL{Prog. Theor. Phys.}{#1}{#2}{#3}}
\nc{\PTPS}[3]   {\JL{Prog. Theor. Phys. Suppl.}{#1}{#2}{#3}}
\nc{\PTEP}[3]   {\JL{Prog. Theor. Exp. Phys.}{#1}{#2}{#3}}
\nc{\PRep}[3]   {\JL{Phys. Rep.}{#1}{#2}{#3}}
\nc{\PPNP}[3]   {\JL{Prog.\ Part.\ Nucl.\ Phys.}{#1}{#2}{#3}}
\nc{\JP}[3]     {\JL{J. of Phys.}{#1}{#2}{#3}}
\nc{\andvol}[3] {{\it ibid.}\JL{}{#1}{#2}{#3}}
%%%%%%%%%%%%%%%%%%%%%%%%%%%%%%%%%%%%%%%%%%%%%%%%%%%%%%%%%%%%%

\end{document}